\UseRawInputEncoding
\documentclass[runningheads]{llncs}
\usepackage{graphicx}
\usepackage{amssymb}
\usepackage{url}
\usepackage{textcomp}
\newcommand{\bhline}[1]{\noalign{\hrule height #1}}

\begin{document}
\title{Source-Free Unsupervised Domain Adaptation with Norm and Shape Constraints for Medical Image Segmentation}
\titlerunning{Source-Free Unsupervised Domain Adaptation}
%
\author{Satoshi Kondo\inst{1}\orcidID{0000-0002-4941-4920}}
%
\authorrunning{S. Kondo}
%
\institute{Muroran Institute of Technology, Hokkaido, Japan\\ 
\email{kondo@mmm.muroran-it.ac.jp}}
%
\maketitle              
\begin{abstract}
Unsupervised domain adaptation (UDA) is one of the key technologies to solve a problem where it is hard to obtain ground truth labels needed for supervised learning. In general, UDA assumes that all samples from source and target domains are available during the training process. However, this is not a realistic assumption under applications where data privacy issues are concerned. To overcome this limitation, UDA without source data, referred to source-free unsupervised domain adaptation (SFUDA) has been recently proposed. Here, we propose a SFUDA method for medical image segmentation. In addition to the entropy minimization method, which is commonly used in UDA, we introduce a loss function for avoiding feature norms in the target domain small and a prior to preserve shape constraints of the target organ. We conduct experiments using datasets including multiple types of source-target domain combinations in order to show the versatility and robustness of our method. We confirm that our method outperforms the state-of-the-art in all datasets.

\end{abstract}

\section{Introduction}

Image segmentation plays a key role in medical image analysis. Deep learning has been widely used for medical image segmentation in recent years. The performance of the segmentation has been much improved with supervised deep learning for a wide variety of imaging modalities and target organs~\cite{Lei20}. However, there are some problems with supervised deep learning for segmentation tasks. The first one is that ground-truth labelling for each pixel is needed in the supervised learning. This is a time-consuming task and it is also difficult to obtain a large number of labeled data for medical images. The second problem is that a model trained using a dataset for a specific modality or organ is hard to use for other modalities or organs since the performance of the model typically degrades.

Unsupervised domain adaptation (UDA) is one of the solutions for these problems~\cite{Kouw21,Toldo20}. UDA is a type of domain adaptation and exploits labeled data from the source domain and unlabeled data from the target one. In general, UDA assumes that all samples from source and target domains are available during the training process. However, this is not realistic assumption under applications where the data privacy issue is concerned, the source and target data come from different clinical sites and so on. To overcome this limitation, UDA without source data, referred to as source-free unsupervised domain adaptation (SFUDA), has been proposed~\cite{Li20}.

SFUDA methods are proposed mainly for image classification tasks~\cite{Kim20,Li20,Liang20,Yang20}. On the other hand, Bateson et al. propose a SFUDA method for a medical image segmentation task~\cite{Bateson20}. This is the only prior art for image segmentation in SFUDA we could find. Their method minimizes a label-free entropy loss defined over target-domain data. However, it is well-known that minimizing this entropy loss alone may result into degenerate trivial solutions, biasing the prediction towards a single dominant class [28]. To overcome this problem, they embed domain-invariant prior knowledge to guide unsupervised entropy training during the adaptation phase, which takes the form of a class-ratio prior. Their method is evaluated using 3D multi-modal magnetic resonance scans of the lower spine.

Our proposed method uses a similar framework with \cite{Bateson20} to address SFUDA image segmentation tasks, but we extend their method in several ways as mentioned below.

As the authors mentioned in a note in \cite{Bateson20}, other estimators than the class ratio could be used in their framework such as region statistics or anatomical prior knowledge. Since the organs to be segmented are the same in both source and target domains in most cases in UDA for medical image segmentation, we think shape prior knowledge must be useful. We introduce a general framework to learn the shape prior knowledge from the source domain data by using an autoencoder network. 

As a general difficulty of domain adaptation tasks, it is pointed out in \cite{Xu19} that the model degradation on the target domain mainly stems from its much smaller feature norms with respect to that of the source domain, for the image classification tasks. To address this issue, we introduce a loss function which constrains the norm of the feature vectors, which is a different approach from the one proposed in \cite{Xu19}.

Besides the technical aspects, the evaluation of domain adaptation methods is conducted by using only one dataset in most papers on domain adaptation in medical image analysis~\cite{Bateson20,Haq20,Liu19,Ouyang19,Varsavsky20}. By contrast, we prepare three different types of source-target combinations for evaluating our method to confirm the versatility and robustness of our method.
We summarize our contributions as follows:

\begin{enumerate}
\item We introduce a prior to preserve shape constraints of the target organ for avoiding the degeneration which happens when using a label-free entropy loss.

\item We introduce a loss function for avoiding feature norms in the target domain small.

\item We compare our method and conventional methods by using datasets including multiple types of source-target domain combinations.

\item We improve the performance of the image segmentation in UDA under a realistic clinical scenario.
\end{enumerate}

\section{Proposed Method}

We address the SFUDA task, which is the UDA task without access to source data in the adaptation phase, for medical image segmentation. We are given $n_{s}$ labeled samples $\{x_{s}^{i},y_{s}^{i}\}_{i=1}^{n_{s}}$from the source domain $D_{s}$ where $x_{s}^{i} \in X_{s}$, $y_{s}^{i} \in Y_{s}$. A source image $x_{s}^{i}$ is $d$-dimensional image and the corresponding ground truth $y_{s}^{i}$ has $K$-class label for each pixel in $x_{s}^{i}$. We are also given $n_{t}$ unlabeled samples $\{x_{t}^{i}\}_{i=1}^{n_{t}}$ from the target domain $D_{t}$ where $x_{t}^{i} \in X_{t}$. The goal of SFUDA for image segmentation is to predict the segmentation labels $\{y_{t}^{i}\}_{i=1}^{n_{t}}$ in the target domain, where $y_{t}^{i} \in Y_{t}$ without access to $\{x_{s}^{i}, y_{s}^{i}\}_{i=1}^{n_{s}}$ in the adaptation phase.

At first, two deep neural networks are trained using the samples from the source domain. The first one is a segmentation network $f$, which is trained using images $\{x_{s}^{i}\}_{i=1}^{n_s}$ and ground truth of the segmentation mask $\{y_{s}^{i}\}_{i=1}^{n_{s}}$. We can use general loss functions for segmentation tasks for training our segmentation network, e.g., we use summation of cross-entropy loss $\mathcal{L}_{c}$ and Dice loss $\mathcal{L}_{d}$ as the basic loss function in our experiments. In~\cite{Xu19}, it is empirically revealed that the erratic discrimination of the target domain mainly stems from its much smaller feature norms with respect to that of the source domain in image classification tasks. From this observation, the authors set up a hypothesis that the domain shift between the source and target domains relies on their misaligned feature-norm expectations. Inspired by this hypothesis, we introduce an additional loss which constrans the norm of the feature vectors. We use the Ring loss~\cite{Zheng18} for our purpose. Assume that the size of the final feature map, i.e., the features just before the final layer of the segmentation network, is [$C, H, W$], where $C$ is the number of channels, $H$ and $W$ are the height and width of the feature map, respectively, the Ring loss is defined as

\begin{equation}
	\mathcal{L}_{c} = \frac{1}{HW}\sum_{h=0}^{H-1}\sum_{w=0}^{W-1} ( \|F(h,w)\|_{2} - R )^{2}, 
\end{equation}

\noindent\\
where $F(h,w)$ is the feature vector at the position $(h,w)$ in the feature map and $R$ is the target value of the norm. In summary, the segmentation network $f$ is trained using the source domain samples by minimizing the following supervised loss $\mathcal{L}_{f} = \mathcal{L}_{c} + w_{d} \mathcal{L}_{d} + w_{r} \mathcal{L}_{r}$, where $w_{d}$ and $w_{r}$ are weights for the Dice and Ring losses, respectively.

%

The second network $g$ is for learning the shape prior. A lot of methods have been proposed for using shape constraints to improve the performance of medical image segmentation~\cite{Bohlender21}. When the target organ is specified, some specific constraints can be applied such as shape compactness or shape smoothness~\cite{Liu20b}. However, we would not like to focus on specific organs. Therefore, we use an autoencoder network for learning shape prior~\cite{Larrazabal19,Oktay18}. The network $g$ is trained using ground truth of the segmentation mask $\{y_{s}^{i}\}_{i=1}^{n_{s}}$ as input and to reconstruct the input segmentation mask. We use summation of cross entropy loss $\mathcal{L}^{'}_{c}$ and Dice loss $\mathcal{L}^{'}_{d}$  as the loss function for the training of the network $g$, since it is important to preserve the shape. The loss function $\mathcal{L}_{g}$ for training the network $g$ is $\mathcal{L}_{g} = \mathcal{L}_{c}' + w_{d}' \mathcal{L}_{d}'$, where $w_{d}'$ is a weight for the Dice loss.

%

In the domain adaptation phase, we use two networks $f$ and $g$ pre-trained using the samples $\{x_{s}^{i}, y_{s}^{i}\}_{i=1}^{n_s}$ in the source domain $\mathcal{D}_{s}$ as mentioned above. The training pipeline in the domain adaptation phase is shown in Fig. 1. We train the segmentation network $f$ with fine-tuning using unlabeled samples $\{x_{t}^{i}\}_{i=1}^{n_t}$ from the target domain $\mathcal{D}_{t}$ and use the network $g$ as trained using the samples in the source domain, i.e., without fine-tuning using the samples in the target domain.

An image from the samples $\{x_{t}^{i}\}_{i=1}^{n_t}$ is inputted to the segmentation network $f$.  The output of the network $f$, which is the predicted segmentation mask, is then inputted to the shape prior network $g$. The output of the network $g$ is the final predicted segmentation mask. From the final segmentation mask, the entropy loss $\mathcal{L}_{e}$ is obtained.

\begin{equation}
	\mathcal{L}_{e} = - \frac{1}{N} \sum_{j=0}^{N-1} \sum_{k=0}^{C-1} p^{j}(k) \log p^{j}(k),
\end{equation}

\noindent\\
where $p^{j}(k)$ is the probability of class $k$ at pixel $j$, $N$ is the number of pixels and $C$ is the number of classes. We also obtain the final feature map from the network $f$ and calculate the Ring loss $\mathcal{L}_{r}$ as in Eq. (1). 

The total loss function $\mathcal{L}_{t}$ is $\mathcal{L}_{t} = \mathcal{L}_{e} + w_{r}' L_{r}$,  where $w_{r}'$ is a weight for the Ring loss.

%
%

\begin{figure}[t]
  \centering
  \includegraphics[scale=0.7]{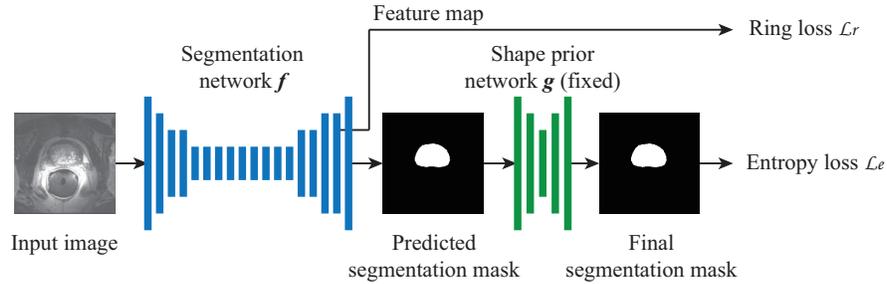}

  \caption{Training pipeline in domain adaptation phase.}
  \label{fig:network}
  \vspace{-.5\baselineskip}
\end{figure}

\section{Experiments}

\subsection{Datasets}

In order to evaluate versatility and robustness of the methods as domain adaptation, we prepared three datasets in wide varieties of source-target domain combinations. We have image (2-dimensional) and volume (3-dimensional) datasets, however we constructed our datasets as the sets of 2-dimentional images and segmentation masks to use the same deep neural network architectures for all datasets. Some specific preprocessing for each dataset was applied when constructing the dataset as mentioned later. At the end of the preprocessing, all images were resized to $256 \times 256$ pixels. All tasks are binary segmentation, i.e., foreground and background classes. Examples of the images in the datasets are shown in Fig. 2. The numbers of samples in source and target domains for each dataset are summarized in Table 1.

\vspace{-\baselineskip}
\subsubsection{CT/MRI Liver.}
This dataset was created from the CHAOS dataset~\cite{Liang20}. The CHAOS dataset provides both abdominal CT and MR data from healthy subjects for single and multiple abdominal organ segmentation. The CT volumes are acquired at the portal venous phase after contrast agent injection. The CT dataset has annotations only for the liver. The MRI dataset includes T1-DUAL in-phase, T1-DUAL oppose-phase and T2-SPIR. The MRI dataset has annotations for the liver, kidney and spleen. We constructed our own dataset using the training dataset in CHAOS which includes 20 CT volumes and 20 MRI volumes. We set a task as the liver segmentation since the CT volumes have annotations only for the liver. Each volume was normalized to zero mean and unit variance as a preprocessing step and the slices including liver region were selected. The CT images and segmentation masks were used as the source domain samples and the MRI T1-DUAL in-phase images were used as the target domain samples. This dataset can be used for domain adaptation between different modalities, i.e., CT and MRI.

\vspace{-\baselineskip}
\subsubsection{MRI Prostate.}
This dataset was created from the Multi-site Dataset for Prostate MRI Segmentation~\cite{Liu20a}. The Multi-site Dataset for Prostate MRI Segmentation is created from NCI-ISBI 2013 dataset~\footnote{https://wiki.cancerimagingarchive.net/display/DOI/NCI-ISBI+2013+Challenge\%3A+Automated+Segmentation+of+Prostate+Structures}, Initiative for Collaborative Computer Vision Benchmarking (I2CVB) dataset~\cite{Lemaitre15} and Prostate MR Image Segmentation 2012 (PROMISE12) dataset~\cite{Litjens14}. The Multi-site Dataset for Prostate MRI Segmentation includes prostate MRI T2W volumes from six sites (sites A to F), in which different imaging conditions, such as the manufacture, field strength (1.5T or 3T), resolution and coil usage, are used. We selected 3T MRI volumes (sites A, C and E), and used volumes from sites A and C (the manufacturer is Siemens) as the source domain and volumes from site B (the manufacturer is GE) as the target domain. Each volume was normalized to zero mean and unit variance as a preprocessing step and the slices including prostate region are selected. This dataset can be used for domain adaptation between different conditions, i.e., manufactures, resolution and coil usage, in the same modality, i.e., 3T MRI T2W. 

\vspace{-\baselineskip}
\subsubsection{Fundus.}
This dataset was created from the Retinal Fundus Glaucoma Challenge (REFUGE) dataset~\footnote{http://refuge.grand-challenge.org/}. The REFUGE dataset consists of retinal fundus images and segmentation masks for optic disc and optical cup regions. The dataset includes 400 training samples with size $2124 \times 2056$ pixels acquired by a Zeiss Visucam 500 camera, and 400 validation samples with size $1634 \times 1634$ pixels collected by a Canon CR-2 camera. We used samples by a Zeiss camera as source domain and the samples by a Canon camera as target domain. As in a similar procedure in~\cite{Liu19}, we cropped a region in each image as the center of optic disc was the center of the cropped area and resized the cropped region to $256 \times 256$ pixels. The sizes of the cropped region were $600 \times 600$ for Zeiss camera and $500 \times 500$ for Canon camera. As for the segmentation mask, we only used the segmentation mask for the optic disc. The images were normalized using mean and standard deviation of ImageNet dataset~\cite{Deng09} since the images in this dataset are RGB color. This dataset can be used for domain adaptation between different manufactures in the same modality, i.e., retinal fundus camera.

\begin{figure}
\begin{minipage}{0.32\hsize}
  \centering
  \includegraphics[width=1.9cm]{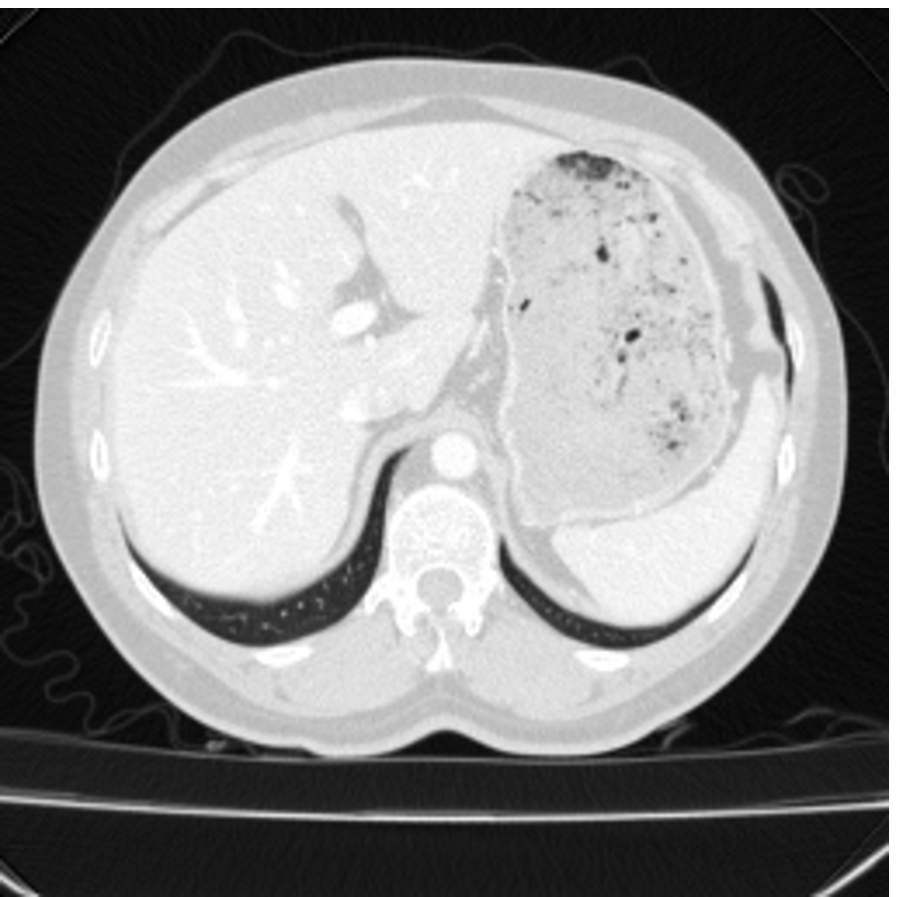}
  \includegraphics[width=1.9cm]{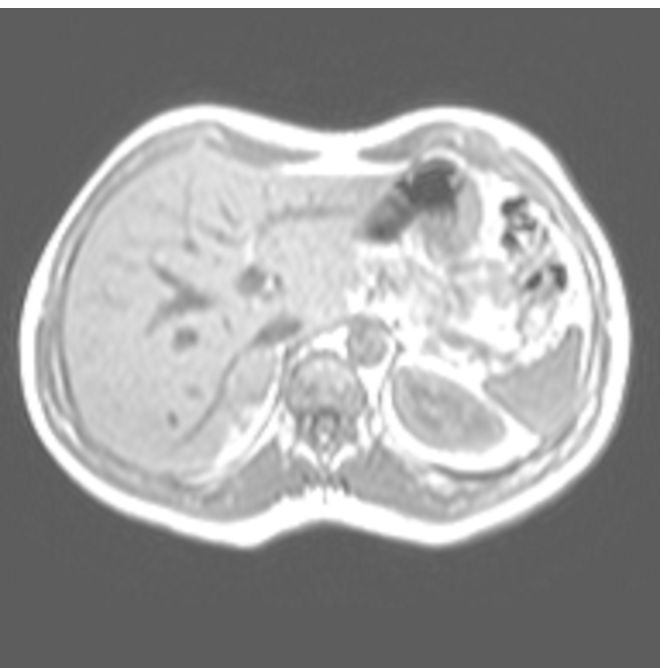}
  (a)
\end{minipage}
\hfill
\begin{minipage}{0.32\hsize}
  \centering
  \includegraphics[width=1.9cm]{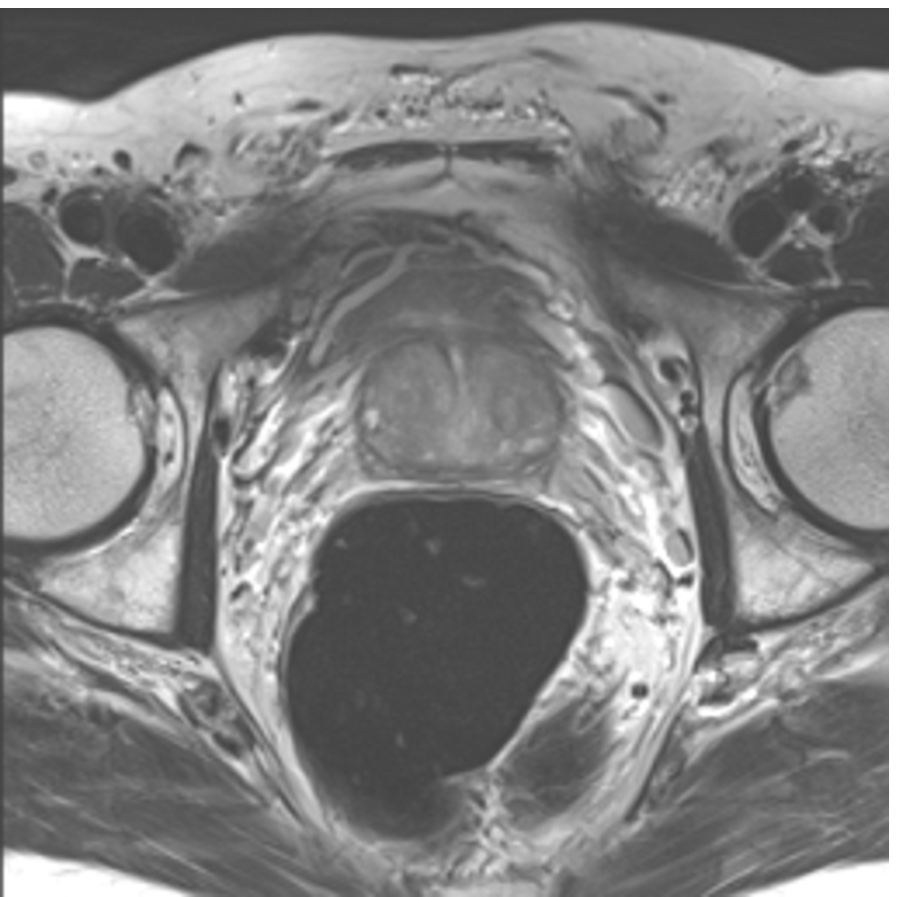}
  \includegraphics[width=1.9cm]{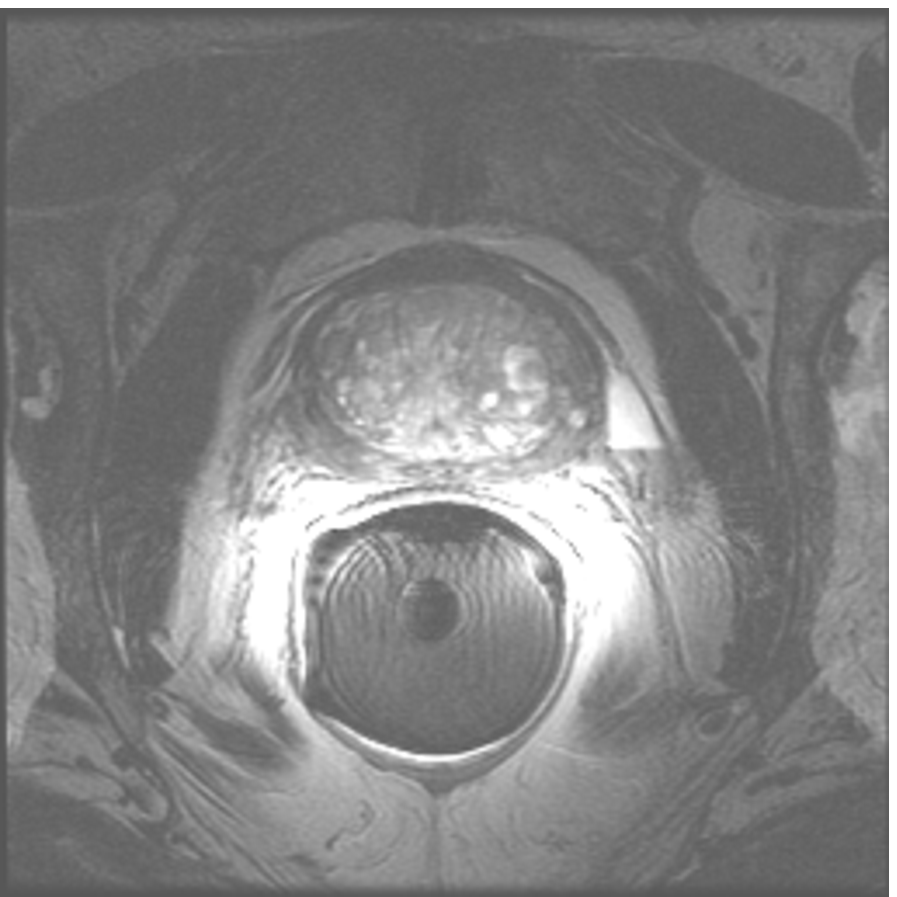}
  (b)
\end{minipage}
\hfill
\begin{minipage}{0.32\hsize}
  \centering
  \includegraphics[width=1.9cm]{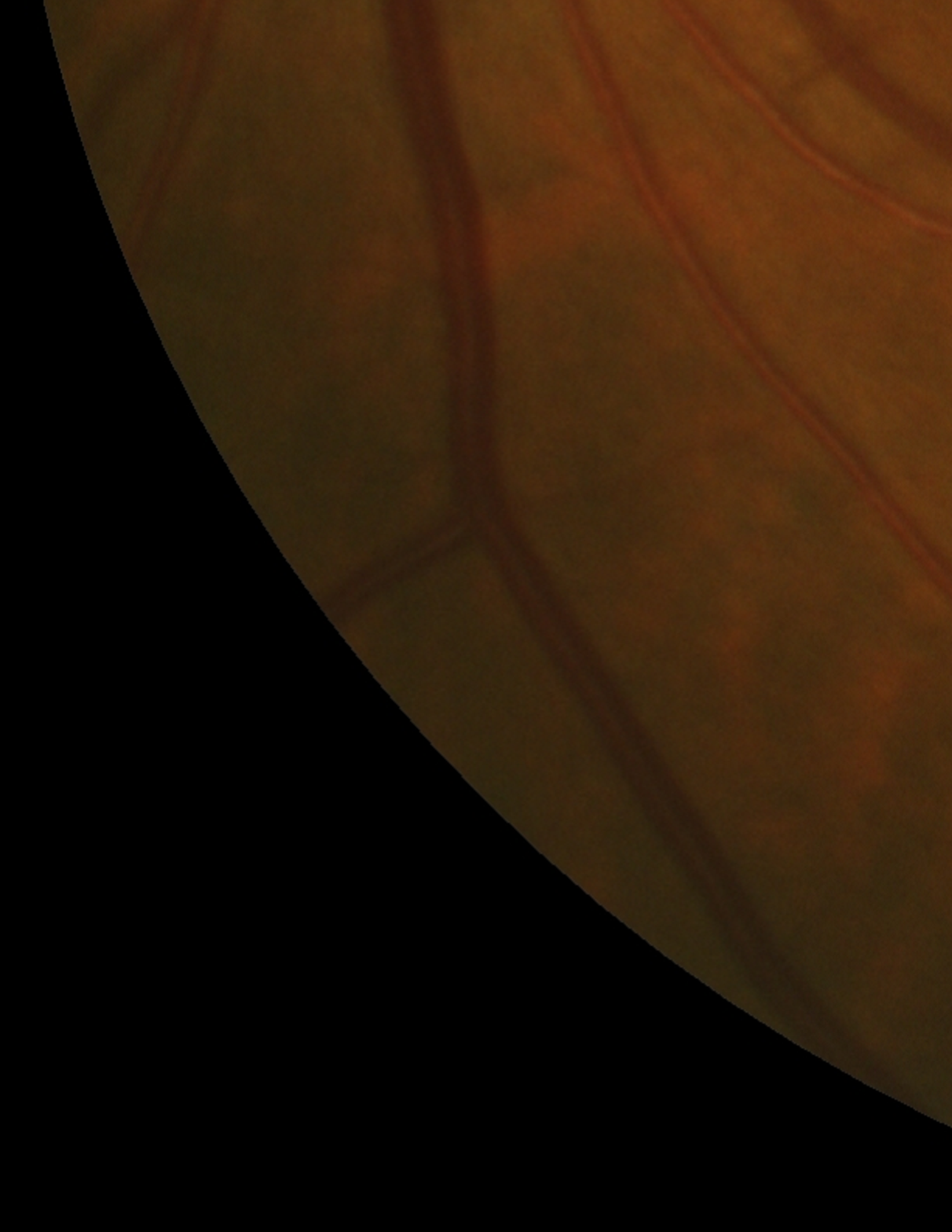}
  \includegraphics[width=1.9cm]{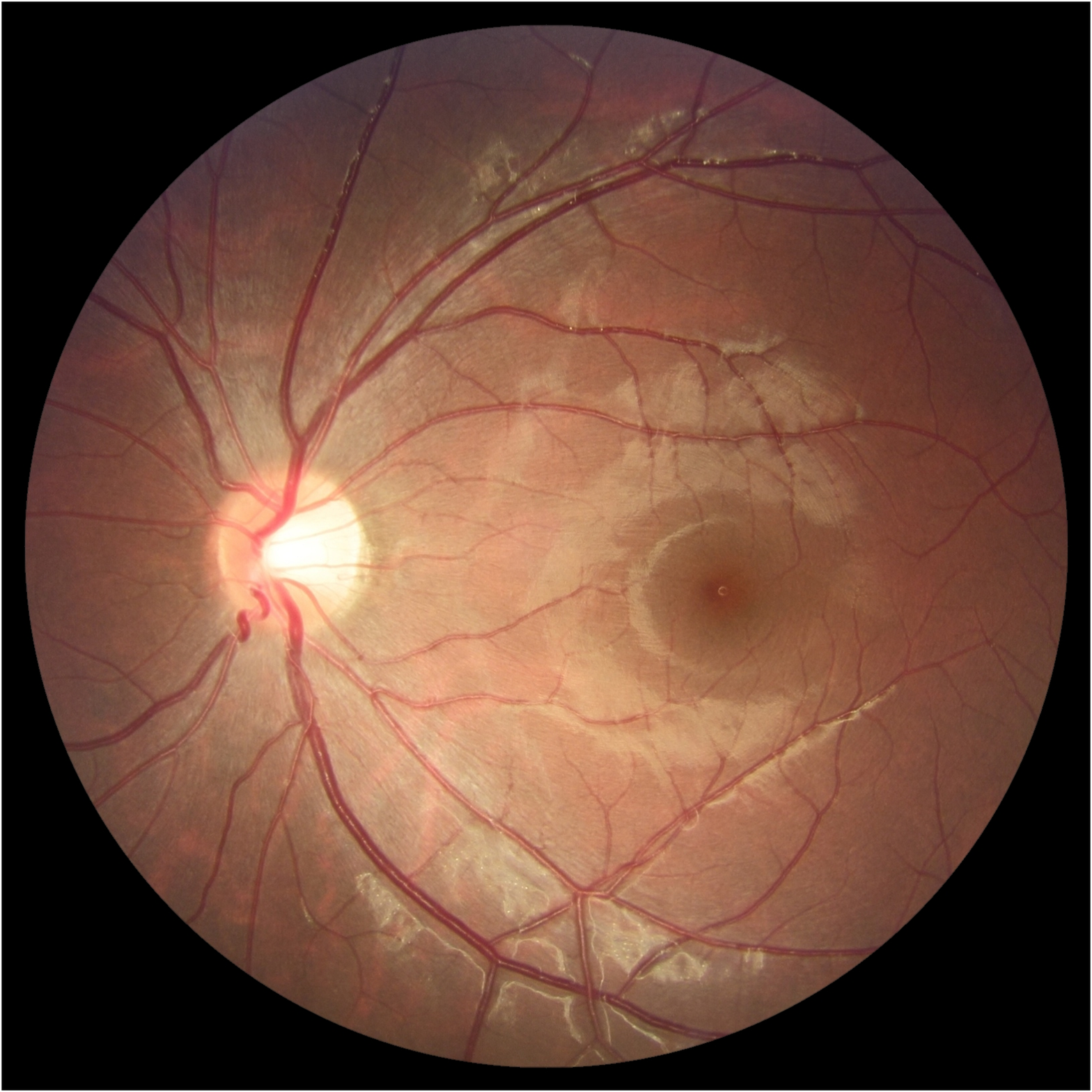}
  (c)
\end{minipage}

\caption{Examples of images in datasets. (a) CT/MRI Liver. (b) MRI Prostate. (c) Fundus. Left and right sides are samples from source and target domains, respectively, in each dataset.}
\vspace{-\baselineskip}
\label{fig:example}
\end{figure}

\subsection{Experimental Conditions}

We employ ENet~\cite{Paszke16} as the segmentation network $f$ and similar autoencoder architecture proposed in~\cite{Larrazabal19} as the shape prior network $g$. These networks are first trained with samples from the source domain as mentioned in Section 2. About eighty percent of the samples in the source domain are used as the training data and the rest of the samples are used as the validation data for each dataset. We employed the model showing the lowest loss value for the validation data for the adaptation phase. The hyperparameters for training the networks are shown in Table 2. These hyperparameters are decided through grid search. As for the other experimental conditions, the parameter $R$ in Eq. (1) is 1.0, the number of epochs is 60, the optimizer is Adam~\cite{Kingma14} and the learning rate changes with cosine annealing. We implemented our method by using Pytorch library~\cite{Paszke19} including conventional methods to compare.

We compared our method with the method proposed in~\cite{Bateson20} (referred to as ``AdaEnt'') as the state-of-the-art for segmentation task in SFUDA. We also evaluated the performance of the segmentation network without DA (referred to as ``No adaptation'') as the lower bound and with supervised learning (finetuning) using ground truth for target domain (referred to as ``Oracle'') as the upper bound. In the domain adaptation phase, the evaluation was conducted with cross-validation of the dataset in the target domain. We divided the dataset into four groups. We used two groups for training, one group for validation and the remaining one group for testing. Thus, we had four folds for cross-validation and the performance was evaluated with average values of the test data in four folds. We selected the model showing the best loss value for the validation data and evaluated the performance of the model for the test data in each fold. We used Dice loss for the evaluation metric for the test data.

We evaluated the performance of our method in the target domain with several different kind of settings, e.g. combination of networks $f$, $g$ and the weights of the loss function $\mathcal{L}_{t}$, as shown in Table 3. We optimized the initial learning rate with grid search for each combination of networks and dataset, including the conventional methods. The optimization was conducted based on the metrics for validation data averaged over four folds. Note that when the Ring loss was not used in the adaptation phase, we used the segmentation network (type 1) shown in Table 2. Otherwise, we used the segmentation network (type 2) shown in Table 2. 

\begin{table}[t]
\begin{center}
\caption{Hyperparameters for deep neural networks in source domain.}\label{tab:parameters}
\begin{tabular}{c|c|c|c|c|c} \bhline{1pt}
\makebox[2.2cm]{Network}
	& \makebox[2.4cm]{
  	   \begin{tabular}{c}
		Initial learning \\ rate 
	   \end{tabular}}
	& \multicolumn{3}{@{}c@{}|}{
		\begin{tabular}{c|c|c}
			\multicolumn{3}{c}{Loss function} \\\hline
			\makebox[2.2cm]{Cross entropy} & \makebox[1.0cm]{Dice} & \makebox[1.0cm]{Ring}
		\end{tabular}}
	& Others \\\bhline{1pt}
\begin{tabular}{c}Segmentation \\ (type 1) \end{tabular}
	& $1.0 \times 10^{-2}$		& \makebox[2.2cm]{\checkmark}	& \makebox[1.0cm]{\checkmark}	& \makebox[1.0cm]{}	
	& \begin{tabular}{c} $w_{d}=1.0$ \\ $(w_{r}=0.0)$ \end{tabular} \\\hline
\begin{tabular}{c}Segmentation \\ (type 2) \end{tabular}
	& $1.0 \times 10^{-2}$		& \checkmark	& \checkmark	& \checkmark
	& \begin{tabular}{c} $w_{d}=1.0$ \\ $w_{r}=1.0$ \end{tabular} \\\hline
Shape prior
	& $2.0 \times 10^{-3}$		& \checkmark	& \checkmark	& 					
	& $w_{d}'=1.0$ \\\bhline{1pt}
\end{tabular}
\end{center}
\end{table}

\begin{table}[t]
\begin{center}
\caption{Settings of our method in domain adaptation phase.}\label{tab:settings}
\begin{tabular}{c|c|c|c} \bhline{1pt}
\makebox[3.2cm]{Name}	& \makebox[2.0cm]{Network $f$} & \makebox[2.0cm]{Network $g$} & \makebox[4.0cm]{Weight in loss function $w_{r}'$} \\\bhline{1pt}
Norm (N) 				& \checkmark	&					& 1.0 \\\hline
Shape (S)				& \checkmark	& \checkmark	& 0.0 \\\hline
Norm + Shape (NS)	& \checkmark	& \checkmark	& 1.0 \\\bhline{1pt}
\end{tabular}
\end{center}
\end{table}

\subsection{Results and Discussion}

Table 4 shows the summary of the experimental results. As shown in Table 4, our method outperformed the state-of-the art (AdaEnt) for all datasets. However, the improvements by our method compared to AdaEnt depended on the datasets, the combination of the networks and the loss functions. Therefore, we will discuss the performance of our method for each dataset separately. 

For the dataset CT/MRI Liver, ``Norm + Shape (NS)'' showed the best performance in our method and the improvement to AdaEnt was 15.8 points in the mean Dice coefficient. On the other hand, ``Shape (S)'' showed poor performance compared to AdaEnt. We think this is because the shapes of livers show significant diversity in 2D slices. The performance of ``Shape'' might improve when we extend our method to 3D volumes.

For the dataset MRI Prostate, ``Shape (S)'' showed the best performance in our method and the improvement to AdaEnt was 4.8 points. ``Norm (N)'' and ``Norm + Shape (NS)'' showed poor performance compared to AdaEnt. We could not identify the reason of the degradation and this is a future item to investigate.

For the dataset Fundus, ``Norm (N)'' showed the best performance in our method and the improvement to AdaEnt was 5.0 points. Note that there were no improvements by AdaEnt to “No adaptation”. In the Fundus dataset, the source and target domain images were captured by different cameras and they had distinct appearances, e.g., color and texture. However, the domain shift was small compared to the other two datasets since the performance difference between ``No adaptation'' and ``Oracle'' was smaller (8.8 points) than others (89.3 and 24.1 points for CT/MRI Liver and MRI Prostate datasets, respectively). Our method improved the performance of UDA even for such a case.

\begin{table}[t]
\begin{center}
\caption{Experimental results. The metric is Dice coefficient. The numbers mean ``average $\pm$ standard deviation''.}\label{tab:result}
\begin{tabular}{c|c|c|c}\bhline{1pt}
\makebox[2.5cm]{Method} 
&
\multicolumn{3}{@{}c@{}}{
\begin{tabular}{c|c|c}
	\multicolumn{3}{@{}c@{}}{Dataset} \\\hline
	\makebox[2.5cm]{CT/MRI Liver} & \makebox[2.5cm]{MRI Prostate} & \makebox[2.5cm]{Fundus} 
\end{tabular}} \\\bhline{1pt}
No adaptation				& \makebox[2.5cm]{1.7 $\pm$ 2.6}	& \makebox[2.5cm]{54.5 $\pm$ 6.5}	& \makebox[2.5cm]{81.0 $\pm$ 0.9} \\\hline
AdaEnt~\cite{Bateson20}	& 47.1 $\pm$ 11.5		& 55.6 $\pm$ 7.1		& 80.7 $\pm$ 1.0 \\\hline
Ours (N)						& 53.5 $\pm$ 9.9		& 50.4 $\pm$ 7.8		& \bf{85.7 $\pm$ 0.8} \\\hline
Ours (S)						& 19.6 $\pm$ 14.1		& \bf{60.4 $\pm$ 5.9}	& 80.8 $\pm$ 0.9 \\\hline
Ours (NS)					& \bf{72.9 $\pm$ 4.6}	& 49.6 $\pm$ 10.4		& 84.0 $\pm$ 1.2 \\\bhline{1pt}
Oracle							& 91.0 $\pm$ 1.4		& 78.6 $\pm$ 6.5		& 89.8 $\pm$ 0.4 \\\bhline{1pt}
\end{tabular}
\end{center}
\end{table}

\section{Conclusions}

We proposed a source-free unsupervised domain adaptation (SFUDA) method for medical image segmentation. In addition to the entropy minimization method commonly used in UDA, we introduced a loss function for avoiding feature norms in the target domain small and a prior to preserve shape constraints of the target organ. We conducted experiments using three datasets including multiple types of source-target domain combinations and confirmed that our method outperformed the state-of-the-art in all datasets.

In the future, our method should be evaluated using more types of source-target domain combinations, datasets having multiple labels and datasets with 3D volumes. In addition, the performance of our method depended on the datasets, the combination of the networks and the loss functions in the experiments. Therefore, we must investigate how we select those conditions only with target samples. 

\vfill


\pagebreak
\bibliographystyle{splncs04}
\bibliography{mybibliography}

\end{document}